\documentclass[10pt,letterpaper,twocolumn]{article} 

\usepackage{ol2}
\usepackage[draft]{hyperref}
\usepackage{amsmath}

\begin{document}

\twocolumn[ 

\title{Ghost imaging lidar via sparsity constraints}


\author{Chengqiang Zhao, Wenlin Gong$^*$, Mingliang Chen, Enrong Li, Hui Wang, Wendong Xu, and Shensheng Han}

\address{Shanghai Institute of Optics and Fine Mechanics,
Chinese Academy of Sciences, Shanghai 201800, China
\\$^*$Corresponding author: gongwl@siom.ac.cn}

\begin{abstract}
For remote sensing, high-resolution imaging techniques are helpful
to catch more characteristic information of the target. We extend
pseudo-thermal light ghost imaging to the area of remote imaging and
propose a ghost imaging lidar system. For the first time, we
demonstrate experimentally that the real-space image of a target at
about 1.0 km range with 20 mm resolution is achieved by ghost
imaging via sparsity constraints (GISC) technique. The characters of
GISC technique compared to the existing lidar systems are also
discussed.
\end{abstract}


 ] 

\noindent

As an important detection tool, lidar has been widely used in remote
sensing in recent decades. Traditional imaging lidar, according to
the measurement modes, can be classified into two types: scanning
imaging lidar and non-scanning imaging lidar (e.g. flash
radiography) \cite{Elachi,Ahola,Anthes}. Scanning imaging lidar
obtains the real-space image of the target by scanning the target
region point-by-point with a pulsed laser \cite{Ahola}. Therefore,
it is difficult to image moving target with high-speed. Non-scanning
imaging lidar \cite{Anthes}, which is characterized by an imaging
system with high resolution and a pulsed flash laser, can cover the
whole field of the target and obtain the target's real-space image
in a single exposure. However, because the light intensity reflected
by the target is divided into many small pixels of charge-coupled
device (CCD) camera, the detection sensitivity is low and the
detection distance of non-scanning imaging lidar is limited by the
signal-to-noise ratio (SNR) of the received photons distributed on
the whole imaging plane and the aperture of the imaging system. In
addition, for scanning imaging lidar the imaging resolution is
limited by the Rayleigh criterion of the emitting aperture
\cite{Zhang1} while by the numerical aperture of the receiving
system for non-scanning imaging lidar.

Recently, there are many researches on the remote sensing with ghost
imaging technology
\cite{Gong,Gong2,Cheng1,Zhang2,Dixon,Hardy,Meyers}. Combined with
sparsity constraints, super-resolution \cite{Donoho,Gong1,Candes},
compressive sensing \cite{Candes,Donoho1,Candes1,Katz,Du,Stern},
compressive radar \cite{Herman} and other techniques
(http://ecos.maths.ed.ac.uk/SPARS11/) are possible. In this Letter,
a practical ghost imaging lidar via sparsity constraints (GISC
lidar) system was proposed and high-resolution imaging was
experimentally demonstrated at a distance about 1.0 km range.

\begin{figure}[htb]
\centerline{\includegraphics[width=8.5cm]{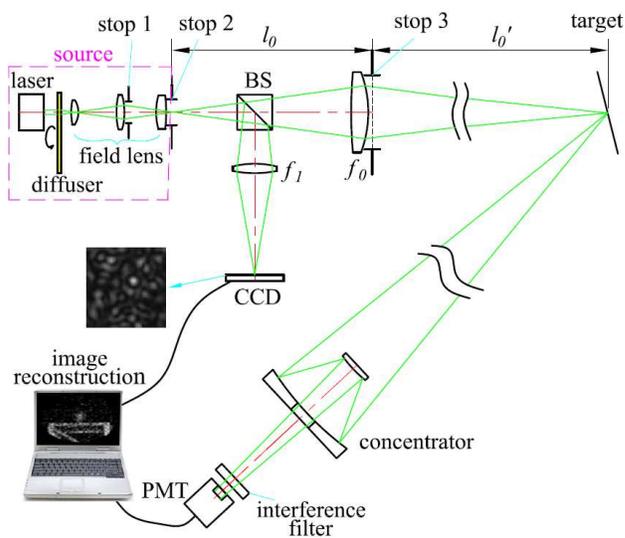}}
\caption{Experimental setup of GISC lidar.}
\end{figure}

Fig. 1 presents the setup of the proposed system. The source, which
consists of a 532 nm coherent solid-state pulsed laser with 10 ns
pulsed width, a rotating diffuser (ground glass disk) and a set of
field lens, forms speckle field at stop 2 (field stop). The light
emitting from the source is divided by a beam splitter (BS) into an
object and a reference paths. In the object path, the light
propagates through the objective lens $f_0$ and then to the target.
The photons reflected by the target are received by a light
concentrator (a Cassegrain telescope with the aperture 420 mm and
the focal length 5 m) and then passes through an interference filter
with 1 nm half-width into a photomultiplier tube (PMT). In the
reference path, the light goes through the reference lens $f_1$ and
then to a CCD camera. In this system, Stop 2 is placed on the
conjugate plane of both the target and the CCD camera, which can
control the field of view (FOV) on the target plane. The field lens
is used to increase the utilization rate of light energy and
generate a visual aperture stop, which is signed as stop 3 in Fig.
1. Therefore, the transverse size of light beam at the objective and
reference lens is controlled by the stop 1, which also ensures that
the entrance pupil is exactly the same for the lens $f_0$ and $f_1$.
In addition, the PMT is used to transform light signals reflected
from the target into electric signals and the interference filter is
used for inhibiting background light.

By exploiting the image's sparsity constraints, CS reconstruction
techniques usually yield, as predicted by theoretical analysis and
confirmed by experiments, better results when the target is sparse
in the representation basis
\cite{Candes,Donoho1,Candes1,Katz,Du,Stern}. The reconstruction of
GISC lidar will be formulated in the CS framework. By reshaping each
of the speckle intensity distribution ($m\times n$ pixels) recorded
by the CCD camera (Fig. 1) into a row vector ($1\times N$,
$N=m\times n$), the measurement matrix ${\rm{A}}$ ($M\times N$) is
obtained after $M$ observations. Meanwhile, the intensities recorded
by the PMT in the object path are arranged as a column vector
${\rm{Y}}$ ($M\times1$). If we reshape the unknown target ($m\times
n$ pixels) into a column vector ${\rm{X}}$ ($N\times1$) and
${\rm{X}}$ can be represented as
${\rm{X}}={\rm{\Psi}}{\rm{\cdot}}{\rm{\ }} \alpha$ such that
$\alpha$ is much sparser (${\rm{\Psi}}$ denotes the transform
operator to the sparse representation basis), then the target's
image can be reconstructed by solving the following convex
optimization program \cite{Gong1,Du,Figueiredo}:
\begin{eqnarray}
{\rm{X}}={\Psi\cdot\rm{\alpha}}; {\rm{ \ }} {\rm{ which \ minimizes:
}}{\rm{ \ }} \frac{{\rm{1}}}{{\rm{2}}}\left\|
{{{\rm{\textbf{Y}}}-\rm{\textbf{AX}}}} \right\|_2^2 + \tau \left\|
{\rm{\alpha}} \right\|_1.
\end{eqnarray}
where $\tau$ is a nonnegative parameter, $\left\| V \right\|_{2}$
and $\left\| V \right\|_{ 1 }$ denote the Euclidean norm and the
$\ell_1$-norm of $V$, respectively.

To experimentally demonstrate the characteristics of GISC lidar, the
concrete parameters of GISC lidar in the experiments are as follows:
the focal length of the lens $f_0$ is 5 m and the magnification of
the lens $f_1$ is 1$\times$. The FOV of the receiving system and the
emitting system at ${l_0}'$=900 m range is about 2 m and 1 m,
respectively. The emitting aperture (stop 3) in the test path is 18
mm (measured value), then the theoretical resolution (Rayleigh
criterion) for traditional imaging system is $\frac{1.22\lambda
{l_0}'}{18}$ mm$\simeq 32.5$ mm. According to the parameters of
emitting system, the pixel size of the CCD camera in the reference
path is set as 27.6 $\mu$m $\times$ 27.6 $\mu$m. The image is
reconstructed using the gradient projection for sparse
reconstruction algorithm \cite{Figueiredo}.

\begin{figure}[htb]
\centerline{\includegraphics[width=8.5cm]{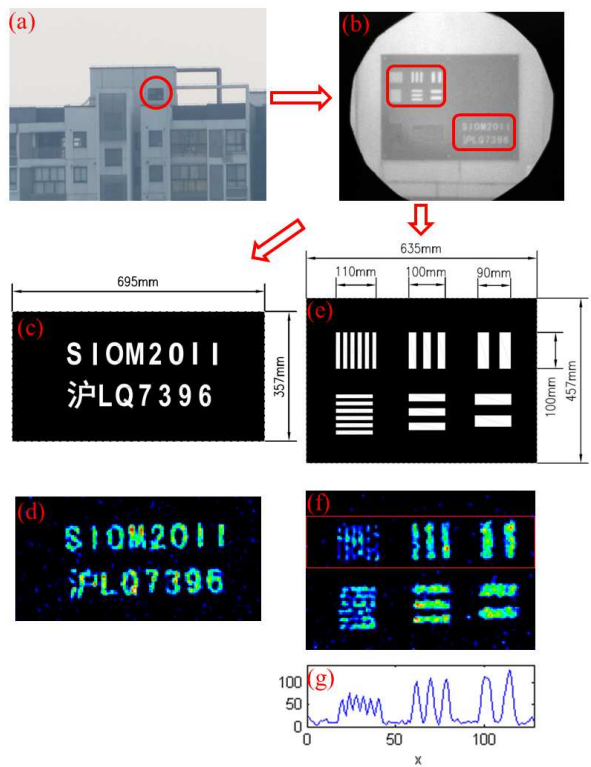}}
\caption{Experimental reconstruction results for high-reflection
targets we proposed at 900 m range (with 3000 measurements). (a, b)
The original target plates imaged by a camera and a telescope,
respectively; (c) the concrete sizes of a standard Chinese vehicle
license plate; (e) the concrete sizes of a set of resolution panels;
(d) and (f) are the targets' images reconstructed by GISC lidar and
the targets are all represented in the space basis; (g) the
cross-section of the rectangular selection box in (f).}
\end{figure}

The first demonstration of GISC lidar was performed using our
designed targets mounted on a building located about ${l_0}'$=900 m
away. The targets are highly reflective: a standard Chinese vehicle
license plate and a set of resolution panels. The line width of the
characters on the vehicle license plate, as shown in Fig. 2(c), is
about 10 mm. The imaging area on the CCD camera is 140$\times$72
pixels, which means that the imaging area on the target plane is 695
mm $\times$ 357 mm. Fig. 2(d) presents the image of the vehicle
license plate reconstructed by GISC lidar. The resolution panels, as
shown in Fig. 2(e), are divided into three groups (six-slit,
three-slit and double-slit) and their center-to-center separation
between slits are 20 mm, 40 mm and 60 mm, respectively. The imaging
area on the CCD camera is 128$\times$92 pixels, which means that the
imaging area on the target plane is 635 mm $\times$ 457 mm. The
reconstructed results of the resolution panels are illustrated in
Fig. 2(f) and Fig. 2(g). From Fig. 2(f) and Fig. 2(g), the
resolution panels with the resolution 20 mm, can be clearly
differentiated, which also demonstrates that super-resolution
imaging is achieved by our GISC lidar.

\begin{figure}[htb]
\centerline{\includegraphics[width=8.5cm]{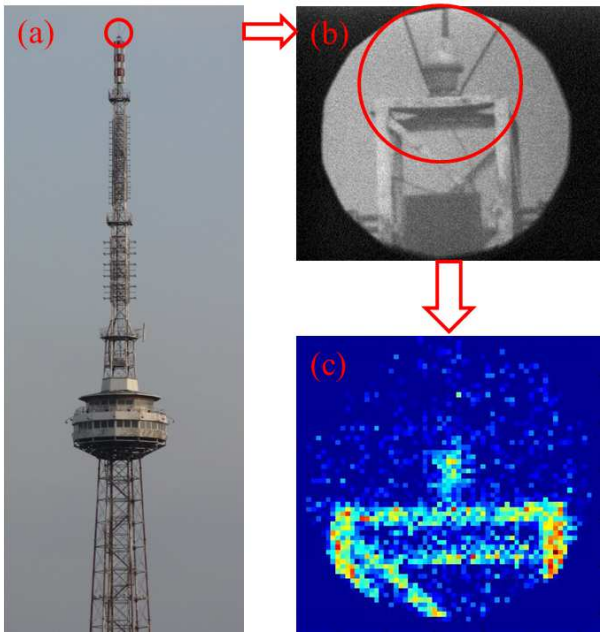}} \caption{
Reconstruction results of a natural target (the overhead target of a
tower) at 720 m range (with 3000 measurements). (a) and (b) are the
image of the target imaged by a camera and a telescope,
respectively; (c) is the target's image reconstructed by GISC lidar
and the target is represented in two-dimensional Discrete Cosine
Transform (2D-DCT) basis.}
\end{figure}

Another demonstration of GISC lidar was conducted to image the
overhead target of a 200 m tower located about ${l_0}'$=720 m away.
The target's images, taken by a camera and a telescope, are shown in
Fig. 3(a) and Fig. 3(b). Fig. 3(c) presents the reconstruction
result of the target (where the target is represented in 2D-DCT
basis). In Fig. 3(c), due to the ultra-low reflectivity and the
specular reflection, the three black antennas are invisible.

Compared with traditional imaging lidar, GISC lidar has both the
advantages of scanning imaging and non-scanned imaging lidar.
Similar to scanning imaging lidar, GISC lidar has high detection
efficiency and long detection distance since all photons collected
by the concentrator illuminate the same PMT. Similar to non-scanning
imaging lidar, the laser pulse emitted form the GISC lidar covers
the whole detection field, therefore, the image can be reconstructed
without scanning the target and the imaging of targets with
high-speed is possible even when the sampling number is far fewer
than the pixel number the desired resolution needed
\cite{Katz,Du,Stern}. In addition, the CCD camera in the reference
path, which limits the sampling speed at present, can be omitted by
using the techniques such as computational ghost imaging
\cite{Shapiro,Bromberg} and encoding the pseudo-thermal light source
\cite{Li}, and the approaches such as image separation
reconstructions \cite{Starck,Kutyniok} are also helpful to the
imaging of targets with high-speed. Furthermore, as indicated by the
results in Fig. 2 and in Ref. \cite{Gong1}, GISC lidar can also
realize supper-resolution imaging.

In conclusion, we experimentally demonstrate a novel imaging lidar
by combining GI method with sparse and redundant representations. We
show that GISC lidar has the advantages of long detection distance,
high imaging speed and super-resolution imaging capabilities.

The work was supported by the Hi-Tech Research and Development
Program of China under Grant Project No. 2011AA120101 and No.
2011AA120102.

\section*{Informational Fourth Page}


\end{document}